%% file: EcConf_ICLR2023v2.tex
\documentclass{article} 
\usepackage{iclr2023_conference,times}

\input{math_commands.tex}

\usepackage{hyperref}
\usepackage{url}
\usepackage{verbatim}
\usepackage{booktabs}
\usepackage{longtable}
\usepackage[ruled,lined]{algorithm2e}
\usepackage{algorithmic}
\usepackage{subfig}
\usepackage{amsmath,amssymb,amsfonts}
\usepackage{caption}
\usepackage{graphicx}
\usepackage{ragged2e} 
\usepackage{booktabs,makecell, multirow, tabularx}
\usepackage{threeparttable}
\UseRawInputEncoding
\usepackage{lipsum}

\newcommand\blfootnote[1]{%
	\begingroup
	\renewcommand\thefootnote{}\footnote{#1}%
	\addtocounter{footnote}{-1}%
	\endgroup
}

\title{EC-Conf: An Ultra-fast Diffusion Model for Molecular Conformation Generation with Equivariant Consistency}

\author{Zhiguang Fan$^{1,2}$, Yuedong Yang$^1$, Mingyuan Xu$^{2*}$, Hongming Chen$^{2*}$  \\
Sun Yat-Sen University,China$^1$\\
Guangzhou National Laboratory, China$^2$\\
\texttt{\{fanzhg5@mail2\}@mail2.sysu.edu.cn} \\
\texttt{\{chen\_hongming\}@gzlab.ac.cn} \\
\texttt{\{mingyuan.xu.sci\}@gmail.com} \\
}

\begin{document}
\blfootnote{*Corresponding Author}  

\maketitle

\begin{abstract}
Despite recent advancement in 3D molecule conformation generation driven by diffusion models, its high computational cost in iterative diffusion/denoising process limits its application. In this paper, an equivariant consistency model (EC-Conf) was proposed as a fast diffusion method for low-energy conformation generation. In EC-Conf, a modified SE (3)-equivariant transformer model was directly used to encode the Cartesian molecular conformations and a highly efficient consistency diffusion process was carried out to generate molecular conformations. It was demonstrated that, with only one sampling step, it can already achieve comparable quality to other diffusion-based models running with thousands denoising steps. Its performance can be further improved with a few more sampling iterations. The performance of EC-Conf is evaluated on both GEOM-QM9 and GEOM-Drugs sets. Our results demonstrate that the efficiency of EC-Conf for learning the distribution of low energy molecular conformation is at least two magnitudes higher than current SOTA diffusion models and could potentially become a useful tool for conformation generation and sampling. We release our code at {https://github.com/DeepLearningPS/EcConf}. 

\end{abstract}

\section{Introduction}

Three dimensional conformations of a molecule can largely influence its biological and physical properties and biological active conformations are usually low-energy conformers (\cite{perola2004conformational}). Thus, many drug design strategies, including structure-based or ligand-based virtual screening, three-dimensional quantitative structure-activity relationships (QSARs) (\cite{cruciani2003three}), and pharmacophore modeling (\cite{schwab2010conformations}), require a fast 3D conformation generation and elaboration protocol for sampling biologically relevant conformations. For the former task, programs such as CONCORD (\cite{hendrickson1993concord}), CORINA (\cite{gasteiger1990automatic}), and OMEGA (\cite{hawkins2010conformer}) are the most popular applications.
These approaches make use of known optimal geometries of molecular fragments that are used as templates for constructing reasonable, low energy 3D models of small molecules. 
These methods usually utilize fallback strategies for structure generation when novel structures appear and belong to rule-based approaches.
However, because producing a single 3D structure of a flexible molecule is almost invariably followed by conformational elaboration, it is the latter process that is both the time and quality bottleneck.

With the advancement of deep learning technologies, deep learning methods have been used to modeling the distribution of 3D bioactive conformation and generate 3D conformer directly. The biggest difference between the above-mentioned traditional 3D conformation generators and deep learning based conformational generative models is that those generative models don$'$t rely on explicit rules to construct 3D conformation but to learn implicitly the distribution of conformational data. Recently, \cite{simm2019generative, xu2021end} employed variational autoencoders (VAEs) (\cite{kingma2013auto}) to predict atomic distances and \cite{dinh2016density} made similar effort with flow-based models. \cite{ganea2021geomol} proposed the GeoMol model, an end-to-end trainable, non-autoregressive conformational generative model consisting of MPNN (\cite{gilmer2017neural}) and self-attention layers (\cite{vaswani2017attention}) to predict both bond lengths and angles. Application of diffusion model-based technology for conformation generation has gained a lot of attention in the latest two years. \cite{shi2021learning} proposed diffusion models by employing denoising score matching (\cite{song2019generative, song2020improved}) to estimate the gradient fields over atomic distances, through which the gradient fields over atomic coordinates can be calculated. They further developed GeoDiff (\cite{xu2022geodiff}) method, which improves the quality of generated conformations model by directly predicting the coordinates without using intermediate distances.

Although various diffusion-based models were proposed for conformation generation, most of them suffered from large iterative diffusion/denoising steps which makes the inference process too slow. Here, we propose an equivariant consistency model (EC-Conf) for fast diffusion of molecular conformation, largely achieving a balance between the conformation quality and computational efficiency. EC-Conf is inspired from the consistency model (\cite{song2023consistency}) based on the probability flow with ordinary differential equation, smoothly transform the conformation distribution into a noise distribution with a tractable trajectory satisfied SE (3)-equivariance. Instead of solving the reverse-time SDE in diffusion models,
 EC-Conf can either directly maps noise vector of prior Gaussian distribution to low-energy conformation in Cartesian coordinates or iteratively maps the solution in the same diffusion trajectory to molecular conformation.
 These characteristics are critical for greatly decreasing the number of iteration in SDE-based diffusion methods while keep the its high capacity to approximate the complex distribution of conformations. The performance of EC-Conf is evaluated on both GEOM-QM9 and GEOM-Drugs datasets. Our model out-perform non-diffusion models and is comparable with SOTA diffusion models, but with at least two magnitudes higher denoising efficiency.

\begin{figure*}[htbp]
	\centerline{\includegraphics[width=5in]{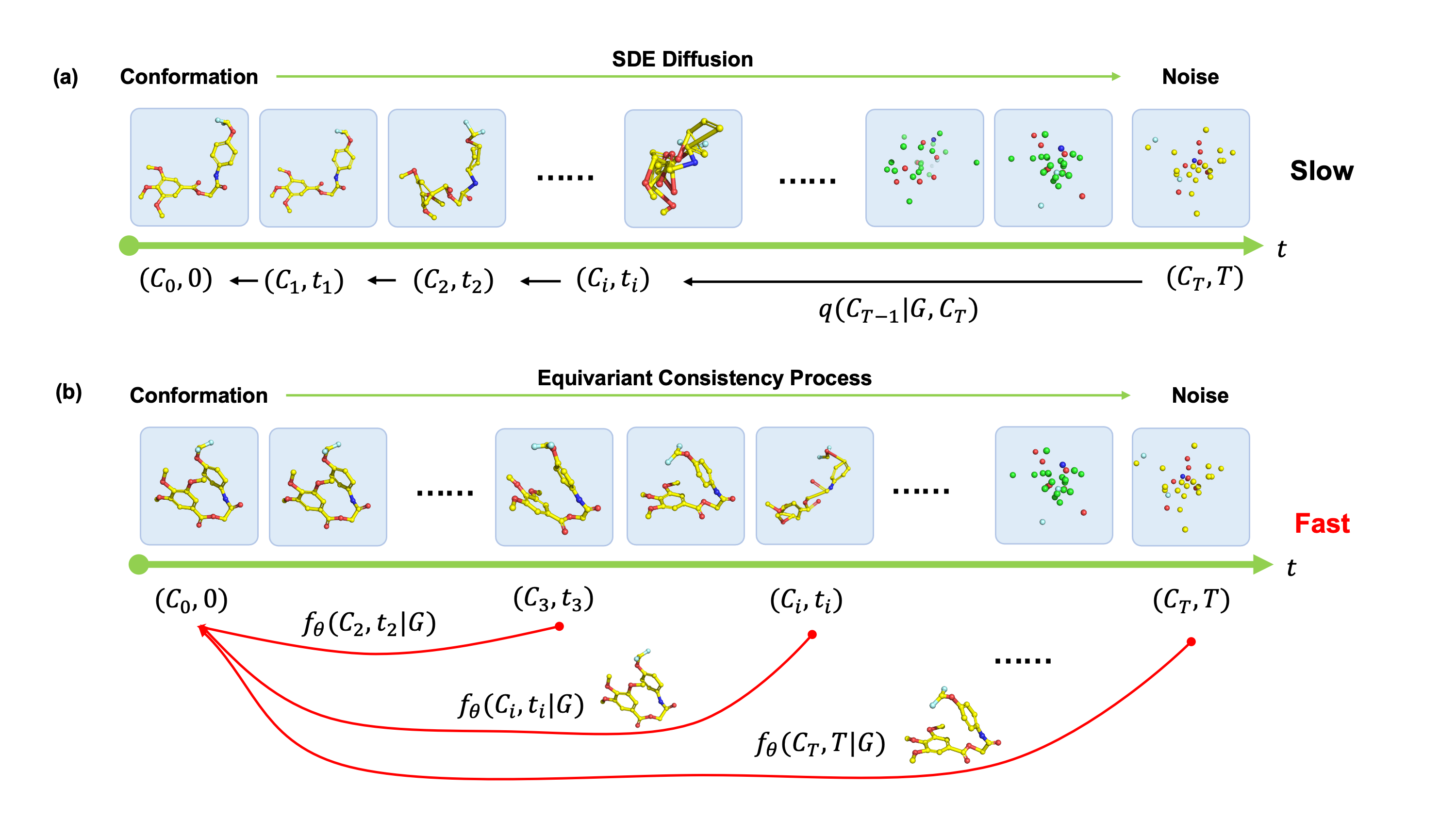}}
	\caption{(a) the diffusion process and the generative phase in normal diffusion models. (b) the equivariant consistency model smoothly transforms the conformation data to the noise among a trackable probability flow, and maps any solution on this trajectory to its origin for fast generation.}
	\label{fig2}
\end{figure*}

\section{Related Work}

\subsection{Deep-learning Based Conformation Generative Models}
With the advancement of deep learning technologies, deep learning methods have been used to directly learn the probability distribution of molecular structures and bring in paradigm change on structure generation. \cite{mansimov2019molecular} reported the early attempt CVGAE to generate 3D conformations in Cartesian coordinates using the variational autoencoder (VAE) architecture in one-shot. However, its performance is not comparable with traditional rule-based methods. \cite{simm2019generative} proposed a conformation generative model based on distance geometry instead of directly modeling the distributions on cartesian coordinates, named Graph-DG. \cite{dinh2016density} proposed GeoMol, which firstly builds the local structure (LS) by predicting the coordinates of non-terminal atoms, and then refine the LS by the predicted distances and dihedral angles. The quality of generated conformation in these one-shot methods has reached the level of traditional methods on small molecules, while there are still improving rooms for conformation generation of drug-like molecules.

Another kind of attempts generate and refine the conformation via a set of sequential samplings instead of one-shot generation. \cite{xu2021learning} proposed CGCF combining the advantages of normalizing flows and energy-based approaches to improve the capacity of estimating the multimodal conformation distribution. Then, \cite{xu2021learning} proposed another score-based method ConfGF reported by \cite{shi2021learning} learns the pseudo-force on each atom via force matching and obtains new conformations via Langevin Markov chain Monte Carlo (MCMC) sampling on the distance geometry (\cite{shi2021learning}). Its performance on the GEOM-Drugs dataset (\cite{axelrod2022geom}) is comparable to that of a rule-based method called experimental-torsion-knowledge distance geometry (ETKDG), which is a conformation generation model implemented in RDKit (\cite{landrum2013rdkit}). \cite{schärfer2013torsion} proposed DMCG trained with a dedicated loss function (\cite{zhu2022direct}). GeoDiff (\cite{xu2022geodiff}) and SDEGen (\cite{zhang2023sdegen}) used a diffusion-based method for conformation generation and also directly predicts the coordinates without using intermediate distances. Similarly, models based on torsion diffusion was also proposed to generate conformations in torsion space instead of cartesian coordinates (\cite{jing2022torsional}). Although these methods improve the quality of generated conformations of drug-like molecules, the slow sampling speed, which due to large number of diffusion iterations, still limit its applications.

\subsection{Diffusions Models}
As mentioned above, various deep generative models have been proposed for conformation generation. These methods could be divided into one-shot model or iterative refinement model. Diffusion-based method belongs to the second class. 

Diffusion model is a family of probabilistic generative model that progressively destruct data by injecting noise, then learn to reverse this process for data generation. It contains three major categories: denoising diffusion probabilistic models (DDPMs) (\cite{ho2020denoising}), score-based generative models (SGMs) (\cite{song2019generative}) and stochastic differential equations (Score SDEs) (\cite{song2020score}). DDPMs and SGMs can be further generalized to the case of infinite time steps or noise levels where the perturbation and denoising processes are solutions to stochastic differential equations (\cite{yang2022diffusion}). Additionally, \cite{song2020score} proved the existence of an ordinary differential equation (ODE), namely probability flow ODE (PF ODE), whose trajectories have the same margin as the reverse-time SDE. The main drawback of diffusion model is its demanding of many iterations to solve the SDE or ODE, which is computationally expensive. For example, GeoDiff, a DDPM model, typically requires 5000 iterations for drug-like molecular conformation generation and SDEGen, a Score SDE based method, requires at least 1500 iterations.

Recently, both learning-free sampling method, including critical damped Langevin diffusion (CLD) (\cite{dockhorn2021score}) and denoising diffusion implicit model (DDIM) (\cite{song2020denoising}), and learning-based sampling method including optimized discretization (\cite{watson2021learning}), knowledge distillation (\cite{salimans2022progressive}) and truncated diffusion methods (\cite{lyu2022accelerating}) are proposed to decrease sampling steps. However, how general of these methods for being applied on molecular conformation generation is still unknown. Another type of method is to reformulating the diffusion process. Recently, \cite{xu2022poisson} proposed Poisson flow generative models (PFGM), treating the data points as charged particles and then transforming the $N$-dimensional target-distribution into a uniform angular distribution in $N+1$ dimensional among the Poisson field. It was demonstrated that PFGMs are 10-20 times faster than diffusion models on image generation tasks. More recently, \cite{song2023consistency} proposed consistency model, a new family of model that generates high quality samples by directly mapping noise to data instead of through the reverse-time SDE. Their results show that consistency models only require a few steps (2-5) of refinement for high quality image generation, which inspired us to incorporate the consistency diffusion process into molecular conformation generation.

\section{Preliminaries}

\subsection{Notations and Problem Definition}

\textbf{Notations.} For a given molecule, it can be represented with an 2D molecular graph $G=\langle\mathcal{V}, \mathcal{E}\rangle$ including features for all nodes $h_v$, $\forall v \in \mathcal{V}$, features for all edges $h_{e_{v, w}}, \forall e_{v, w} \in \mathcal{E}$ and a set of coordinates $C=\left\{c_1, c_2, \ldots c_v\right\}$ of atoms, $\forall v \in \mathcal{V}$.

\textbf{Problem Definition.} Molecular conformation generation is defined as a conditional generative problem with a given 2D molecular graph $G$, where we only focus on the low-energy conformations $C$. For provided multiple molecular graphs $\mathcal{G}$ , we aim to learn a transferable approximate generative model $P_\theta(C \mid G)$ mapping the Boltzmann distribution of $C$ under condition of $G \in \mathcal{G}$ to a prior Gaussian distribution, which conformations can be sampled for a given $G$.

\subsection{Equivariance}
For a molecule conformation, its coordinates are allowed to be changed via translation and rotation in 3D space, while its scalar properties should still be invariant e.g., energy or ADMET properties. This feature is also called SE (3) equivariance, which should be satisfied for 3D based molecular generative models.  Formally, a function:$f: x \rightarrow y$ is equivariant to a group of transformation $\mathbb{G}$, thus
\begin{equation}
	f\left(D_x(g) x\right)=D_y(g) f(x), g \in \mathbb{G}
	\label{eq1}
\end{equation}
where $D_x(g)$ and $D_y(g)$ are transformation matrices parametrized by $g$ in $\mathcal{X}$ and $\mathcal{Y}$. Our EC-Conf models satisfied this SE (3) equivariance to make the estimated likelihood of conformations unaffected by the translation and rotation operation.

\section{EcConf Method}
\subsection{Background}
Here, we elaborate on the proposed equivariant consistency framework for conformation generation. In DDPM-based diffusion process like GeoDiff, noise from fixed posterior distributions $q\left(C^t \mid C^{t-1}\right)$ is gradually added until the ground truth conformation $C^0$ is completely destroyed with $T$ time steps. During generation process, an initial state $C^T$ is sampled from standard Gaussian distribution, the conformation is progressively refined via the model learned Markov kernels $p_\theta\left(C^{t-1} \mid G, C^t\right)$ for a given molecular graph $G$. \cite{song2020score} have demonstrated that this diffusion process can be described as a discretization process on the time and noise in form of stochastic differential equation (SDE) as defined in equation \ref{eq2}.
\begin{equation}
	d C=f(C, t) d t+g(t) d w
	\label{eq2}
\end{equation}
This process can be reversed by solving the following reverse-time SDE:
\begin{equation}
	d C=\left[f(C, t)-g(t)^2 \nabla_C \log p_t(C)\right] d t+g(t) d \bar{w}
	\label{eq3}
\end{equation}

Where $f(C, t)$ and $g(t)$ are diffusion and drift functions of the SDE,  $w$ and $\bar{w}$ are the standard Brownian motion when time flows forward and backwards, respectively, and $\nabla_C \log p_t(C)$ is the gradient of the log probability density. The denoising process can also be described in the form of ordinary differential equation (ODE), representing the probability flow to the same marginals as the reverse-time SDE, i.e. PF ODE (\cite{song2020score}).
\begin{equation}
	d C=\left[f(C, t)-1 / 2 g(t)^2 \nabla_C \log p_t(C)\right] d t
	\label{eq4}
\end{equation}
\cite{karras2022elucidating} further simplified the equation \ref{eq4} by setting $f(C, t)=0$ and $g(t)=\sqrt{2 t}$. In this case, the $p_t(C)=p_{\text {data }}(C) \otimes \mathcal{N}\left(0, T^2 I\right)$  and get the empirical PF ODE as shown in equation \ref{eq5}.
\begin{equation}
	\frac{d C_t}{d t}=-t s_\phi\left(C_t, t\right)
	\label{eq5}
\end{equation}

Which allows us to sample $\widehat{C}_T$ from $\pi=\mathcal{N}\left(0, T^2 I\right)$ to initialize the empirical PF ODE and solve it backwards in time with any numerical ODE solver including Euler and Heun solvers. Then we get a solution trajectory $\left\{C_t\right\}_{t \in[0, T]}$ and the $\widehat{C}_0$ is an approximate sample from the conformation distribution.

\subsection{Definition of EC-Conf}
Given a solution trajectory  $\left\{C_t\right\}_{t \in[\epsilon, T]}$ of the ODE in equation \ref{eq5}, where $\epsilon$ is an approximate to zero. Different from the diffusion model as discussed above, we desire an equivariant consistency function $f:\left(C_t, t \mid G\right) \rightarrow C_\epsilon$ which satisfies both a boundary condition and the self-consistency. The boundary condition means that there exists a given function value $f\left(C_\epsilon, \epsilon\right)=C_\epsilon$ at the small time step $\epsilon$, while self-consistency means that the output of function $f\left(C_t, t \mid G\right)$ is consistent for arbitrary pairs of $\left(C_t, t\right)$ on the same ODE trajectory, i.e., $f\left(C_t, t \mid G\right)=f\left(C_{t^{\prime}}, t^{\prime} \mid G\right)$ for all $t, t^{\prime} \in[\epsilon, T]$. Once both of these two conditions are satisfied, all the solution $\left\{C_t\right\}_{t \in[\epsilon, T]}$ on the ODE trajectory can directly be mapped to the original ground truth $C_0$. As proved in \cite{song2023consistency}, the transformation between $p_{\text {data }}(C \mid G)$ to $\pi \sim \mathcal{N}\left(0, T^2 I\right)$ were built. Additionally, the consistency function should also be SE (3)-equivariant, which means it need to satisfy equation \ref{eq1}. These characteristics not only allow EC-Conf model to make one step generation from the prior distribution but also can improve the quality of generation by chaining the outputs of consistency models at multiple time steps as shown in Figure \ref{fig2} (b).

We parameterize the consistency model with a learnable function $f_\theta$ using skip connections as shown in equations \ref{eq6}, \ref{eq7} and \ref{eq8}.
\begin{equation}
	f_\theta(C, t \mid G)=c_{\text {skip }}(t) C+c_{\text {out }}(t) F_\theta(C, t \mid G)
	\label{eq6}
\end{equation}

\begin{equation}
	c_{\text {skip }}(t)=\frac{\sigma_{\text {data }}^2}{(t-\epsilon)^2+\sigma_{\text {data }}^2}
	\label{eq7}
\end{equation}

\begin{equation}
	c_{\text {out }}(t)=\frac{\sigma_{\text {data }}(t-\epsilon)}{\sqrt{\sigma_{\text {data }}^2+t^2}}
	\label{eq8}
\end{equation}

Where $F_\theta$ is a $(G, t)$ related SE (3)-equivariance model to make $f_\theta$ satisfied the equivariance. If $c_{\text {skip }}(\epsilon)=1$ and $c_{\text {out }}(\epsilon)=0$ for $t=\epsilon$, make $f_\theta(\epsilon)=C_\epsilon$ satisfied the boundary conditions naturally, and both $c_{\text {skip }}, c_{\text {out }}, F_\theta$ are differentiable for training continuous-time consistency model. 

\begin{figure*}[htbp]
	\centering
	\centerline{\includegraphics[width=5in]{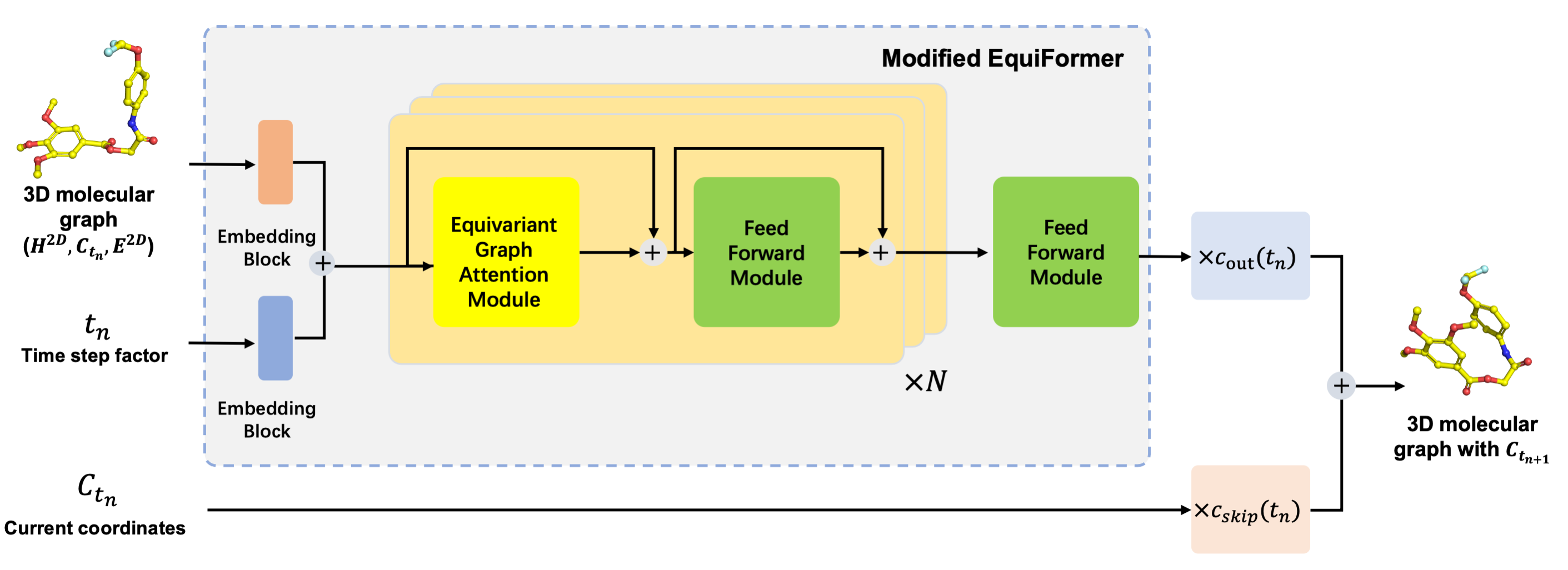}}
	\caption{The model architecture in EC-Conf.}
	\label{fig3}
\end{figure*}

\subsection{Model Architecture}
In principle, EC-Conf, as a generative learning frame work, could adopt any type of Graph conditioned SE (3)-equivariance neural network for modelling $F_\theta$ and is then plugined into the equation \ref{eq6}. Here, we employed and modified the Equiformer \cite{liao2022equiformer}, a SE (3)-equivariant transformer model based on the irreducible representations and depth-wise tensor production based equivariant attention mechanisms, by incorporating embedding of time step factor, to model the $F_\theta$, and the overall architecture of EC-Conf is shown in Figure \ref{fig3}. The time factor $t_n$ and the atom type are embedded with a linear layer respectively and added together as the input of EquiFormer. As the time factor $t_n$ and atom type are both invariant with coordinate systems, thus, this modification won$'$t affect the SE (3)-Equivariance of Equiformer. 

Due to the $C_t$ and $F_\theta$ are both SE (3) equivariant, and $c_{\text {skip }}(t)$ and $c_{\text {out }}(t)$ are invariant scalar with given , the SE (3)-equivariance are guaranteed here. 

\IncMargin{1em}
\begin{algorithm}
	\footnotesize
	\SetKwData{Left}{left}\SetKwData{This}{this}\SetKwData{Up}{up}
	\SetKwFunction{Union}{Union}\SetKwFunction{FindCompress}{FindCompress}
	\SetKwInOut{Input}{input}\SetKwInOut{Output}{output}
	\Input{ dataset $\mathcal{D}=\left\{G_i, C_i\right\}_{i \in M}$, where $M$  refer to the conformation number in $\mathcal{D}$, initial model parameter ${\theta}$, learning
		rate $\eta$, step schedule $N(\cdot)$, EMA decay rate schedule 
		$\mu(\cdot)$,
		${\theta}^{-} \leftarrow {\theta}$ and $k \leftarrow 0$;\newline
	}

	\BlankLine
	\Repeat { \bf{convergence}}
	{\ \ 
		\bf{Sample} $\quad G, C \sim \mathcal{D}$, and $n \sim \mathcal{U}[1, N(k)-1]$ ;\newline
		\bf{Sample} $\quad {z} \sim \mathcal{N}({0}, {I})$ ;\newline
		$\mathcal{L}\left({\theta}, {\theta^{-}}\right) \leftarrow MSE \left(f_{{\theta}}\left(C+t_{n+1} \cdot z, t_{n+1} \mid G\right), f_{{\theta^{-}}}\left(C+t_n \cdot z, t_n \mid G\right)\right)$ ;\newline
		${\theta} \leftarrow {\theta}-\eta \nabla_{{\theta}} \mathcal{L}\left({\theta}, {\theta}^{-}\right)$ ;\newline
		${\theta^{-}} \leftarrow \mu(k) {\theta^{-}}+(1-\mu(k)) {\theta}$ ;\newline
		$k \leftarrow k+1$ ;\newline}

	\caption{Equivariant Consistency Training}
	\label{algorithm1}
\end{algorithm}
\DecMargin{1em}

\subsection{Model Training and Conformation generation}
In order to achieve self-consistency in the diffusion model, the training object is to minimize the difference between two adjacent time steps, i.e., $f_\theta\left(C_{n+1}, t_{n+1} \mid G\right)$ and  $f_\theta\left(C_n, t_n \mid G\right)$. As shown in Algorithm \ref{algorithm1}, for a conformation $C$ sampled from the dataset, we use $C+t_{n+1} \cdot z$ and $C+t_{n} \cdot z$ to replace $C_{n+1}$, $C_n$ on the ODE trajectory $\left\{C_t\right\}_{t \in[\epsilon, T]}$,  thus the $f_\theta$ can be trained with the following:
\begin{equation}
	\mathcal{L}(\theta)=MSE\left(f_\theta\left(C+t_{n+1} \cdot z, t_{n+1} \mid G\right), f_\theta\left(C+t_n \cdot z, t_n \mid G\right)\right)
\end{equation}

To make the training process more stable and improve the final performance of $f_\theta$, another function $f_{\theta^{-}}$  with the learnable parameters $\theta^{-}$, which is the exponential moving average (EMA) of the parameter $\theta$ of original function $f_\theta$ during training process, is introduced and eventually the difference between $f_\theta\left(C+t_{n+1} \cdot z, t_{n+1} \mid G\right)$ and $f_{\theta^{-}}\left(C+t_n \cdot z, t_n \mid G\right)$ is minimized. Here, we refer $f_\theta$ as  "online network" and the $f_{\theta^{-}}$ as the "target network" as mentioned in \cite{song2020score} ’s work. The loss function is reformulated as follows:
\begin{equation}
	\mathcal{L}\left(\theta, \theta^{-}\right)=MSE\left(f_\theta\left(C+t_{n+1} \cdot z, t_{n+1} \mid G\right), f_{\theta^{-}}\left(C+t_n \cdot z, t_n \mid G\right)\right)
\end{equation}

Parameter $\theta$ is updated with stochastic gradient descent, wile $\theta^{-}$ is updated with exponential moving average as shown in equation \ref{eq11}, where $\mu$ is the decay rate predefined by EMA schedule.
\begin{equation}
	\theta^{-}=\mu \theta^{-}+(1-\mu) \theta
	\label{eq11}
\end{equation}
Bying doing in this way, we could perform the equivariant consistency training to get the approximate function $f_\theta$ to $f$, namely equivariant consistency conformation generative model (EC-Conf).

In the conformation generation phase, samples are drawn from the initial distribution $\widehat{C}_T \sim N\left(0, T^2 I\right)$ and the consistency model is used to generate conformations: $\widehat{C}_\epsilon=f_\theta\left(C_T, T\right)$. The conformers are refined with greedy algorithm by alternating denoising and noise injection steps with a set of time points $\left\{\tau_1, \tau_2 \ldots \tau_{N-1}\right\}$ as shown in Algorithm \ref{algorithm2}.

\IncMargin{1em}
\begin{algorithm}
	\footnotesize
	\SetKwData{Left}{left}\SetKwData{This}{this}\SetKwData{Up}{up}
	\SetKwFunction{Union}{Union}\SetKwFunction{FindCompress}{FindCompress}
	\SetKwInOut{Input}{input}\SetKwInOut{Output}{output}
	\Input{Consistency model ${f}_{{\theta}}(\cdot, \cdot)$,
		sequence of time points $\left\{\tau_1, \tau_2 \ldots \tau_{N-1}\right\}$ that $\tau_1>\tau_2>\cdots>\tau_{N-1}$,
		initial noise $\widehat{C}_T$ sampled from $\mathcal{N}\left(0, T^2 I\right)$,
		{$C \leftarrow f_\theta\left(\widehat{C}_T, T \mid G\right)$};\newline
	}

	\BlankLine
	\For{$n=1$ \bf{to} $N-1$}{
		\bf{Sample} $\quad {z} \sim \mathcal{N}({0}, {I})$ ;\newline
		\emph{$\widehat{C}_{\tau_n} \leftarrow C+\sqrt{\tau_n^2-\epsilon^2} \cdot z$}\;
		\emph{$C \leftarrow f_\theta\left(\widehat{C}_{\tau_n}, \tau_n \mid G\right)$}\;
	}
	
	\Output{{${C}$}}\

	\caption{Equivariant Consistency Sampling}
	\label{algorithm2}
\end{algorithm}
\DecMargin{1em}

\section{Experiment}
This section presents an empirical performance evaluation of EC-Conf on the task of 3D conformation generation for small and drug-like molecules.

\subsection{Experiment Setup}

Following previous work, we also use GEOM-QM9 and GEOM-Drugs datasets for evaluation (\cite{liao2022equiformer}). To make a fair benchmark study, we used the same training, validation and test sets produced by \cite{shi2021learning} on both datasets. The GEOM-QM9 is split into training, validation and test set of 39860, 4979, 200 unique molecules, corresponding to 198637, 24825 and 24800 conformations. The GEOM-Drugs set contains 39852, 4983 and 200 molecules in training, validation and test set, corresponding to 198873, 24875 and 14299 conformations, respectively. 
We examined model performance on the same test set used by GeoDiff model.
In current study, following hyperparameters $\sigma_{\text {data }}=0.5, \epsilon=10^{-8}, T=80$ are used. For a given molecule, in case there are $K$ ground truth conformations in test sets, $2K$ conformations are sampled for evaluation.

\subsubsection{Baselines}
In our benchmark, 8 recent or established SOTA models including GraphDG (\cite{simm2019generative}), ConfVAE (\cite{DBLP:conf/icml/XuWLSBGT21}), CGCF (\cite{xu2021learning}), GeoMol (\cite{ganea2021geomol}), ConfGF \cite{shi2021learning}, SDE-Gen (\cite{zhang2023sdegen}), GeoDiff (\cite{xu2022geodiff}) and ETKDG (\cite{landrum2013rdkit}) methods in RDKit. The results of GraphDG, ConfVAE, CGCF, GeoMol, GonfGF, GeoDiff and ETKDG are borrowed from \cite{xu2022geodiff}, while the performance of SDEGen (\cite{zhang2023sdegen}) are evaluated on the same 200 molecules with the provided pre-trained models and different settings by ourselves.

\subsubsection{Evaluation Metrics}

This task aims to measure the quality and diversity of conformations generated by different models. We follow \cite{ganea2021geomol} in evaluating four metrics built on the root mean square deviation (RMSD), defined as the normalized Frobenius norm of two atomic coordinate matrices, after alignment by the Kabsch algorithm (\cite{kabsch1976solution}). Formally, let $S_g$ and $S_r$ denote the set of generated conformations and the set of reference conformations, respectively, then the coverage and matching measures (\cite{xu2021learning}) following the traditional recall measure can be defined as follows:
\begin{equation}
	\operatorname{COV}-\mathrm{R}\left(S_g, S_r\right)=\frac{1}{\left|S_r\right|}\left|\left\{\mathcal{C} \in S_r \mid \operatorname{RMSD}(\mathcal{C}, \hat{\mathcal{C}}) \leq \delta, \hat{\mathcal{C}} \in S_g\right\}\right|,
\end{equation}

\begin{equation}
	\operatorname{MAT-R}\left(S_g, S_r\right)=\frac{1}{\left|S_r\right|} \sum_{\mathcal{C} \in S_r} \min _{\hat{\mathcal{C}} \in S_g} \operatorname{RMSD}(\mathcal{C}, \hat{\mathcal{C}})
\end{equation}

Where $\delta$ is a predefined threshold. The other two metrics, COV-P and MAT-P, are inspired by precision and can be defined similarly, but with the generated and referenced set exchanged. In practice, the $S_g$ of each molecule is set to twice the size of the $S_r$. Intuitively, the COV score measures the percentage of structures in one set that are covered by another set, where coverage means that the RMSD between two constructs is within a certain threshold $\delta$. In contrast, the MAT score measures the average RMSD of one set of conformations against another set of nearest neighbors. In general, a higher COV rate or a lower MAT score indicates that a more realistic conformation is generated. Moreover, the precision metric relies more on quality, while the recall metric focuses more on diversity. Given the specific scenario, either metric may be more attractive. 
Following previous works (\cite{xu2021learning, ganea2021geomol}), $\delta$ is set as $0.5 \AA$ and $1.25 \AA$ for GEMO-QM9 and GEMO-Drugs datasets respectively.

\subsection{Results and Discussion}

\subsubsection{The performance of EC-Conf with different training time steps}
The first thing to investigate is how training time points influence the model performance. Various models were trained by using different training time steps and evaluated on a random  test set. For EC-Conf, user can set the maximal time points in both forward and reverse, corresponding to the training and generation phase. We first evaluated the performance of EC-Conf trained with different maximal time points of 5, 10, 15, 25, 50 in forward ODE process, the iteration steps during generation are the same as training phase and the results are as shown in Table \ref{tab1} of Appendix \ref{app1}. It seems that both COV-R and MAT-R metrics reached best level when the training time point was set to 25. The performance on COV-R and MAT-R got improved when the training time steps increased from 2 to 25, while it got worse for diffusion step of 50 , which is probably due to the error accumulation during the sampling. For COV-P and MAT-P, the performance of EC-Conf reached the top when training time point was set to 15, and got worse for number of iteration larger than 15. Take both Recall and Precision measurement into consideration, it seems that the models with diffusion steps of 25 in the forward ODE trajectories gave best result in general and it was set as optimum diffusion step in training for the following experiments.

\begin{table}[htbp]
	\centering
	\scriptsize
	\caption{The benchmark results on GEOM-Drugs test set}
	\begin{threeparttable}
		\begin{tabular}{l|c|cccc|cccc}
			\toprule
			\textbf{Dataset} &   &      \multicolumn{4}{c|}{\textbf{Drugs}} & \multicolumn{4}{c}{\textbf{Drugs}} \\
			\midrule
			\textbf{Task} &    &     \multicolumn{4}{c|}{\textbf{Conformation Generation}} & \multicolumn{4}{c}{\textbf{Conformation Generation}} \\
			\midrule
			\multicolumn{1}{l|}{\textbf{Metric} $\rightarrow$} & \multirow{2}[2]{*}{\textbf{Steps}} &       \multicolumn{2}{c}{COV-R $\uparrow$} & \multicolumn{2}{c|}{MAT-R $\downarrow$} & \multicolumn{2}{c}{COV-P $\uparrow$} & \multicolumn{2}{c}{MAT-P $\downarrow$} \\
			\multicolumn{1}{l|}{\textbf{Model $\downarrow$}} &  &   Mean  & Median & Mean  & Median & Mean  & Median & Mean  & Median \\
			\midrule
			RDKIT & 1     &        0.6091  & 0.6570  & 1.2026  & 1.1252  & \textbf{0.7222 } & \textbf{0.8872 } & \textbf{1.0976 } & \textbf{0.9539 } \\
			GraphDG & 1     &        0.0827  & 0.0000  & 1.9722  & 1.9845  & 0.0208  & 0.0000  & 2.4340  & 2.4100  \\
			ConfVAE & 1     &        0.5520  & 0.5943  & 1.2380  & 1.1417  & 0.2296  & 0.1405  & 1.8287  & 1.8159  \\
			Geomol & 1     &        \textbf{0.6716 } & \textbf{0.7171 } & \textbf{1.0875 } & \textbf{1.0586 } & $-$      & $-$      & $-$     & $-$  \\
			\midrule
			CGCF  & 1000  &       0.5396  & 0.5706  & 1.2487  & 1.2247  & 0.2168  & 0.1372  & 1.8571  & 1.8066  \\
			ConfGF & 5000 &       0.6215  & 0.7093  & 1.1629  & 1.1596  & 0.2342  & 0.1552  & 1.7219  & 1.6863  \\
			SDEGen\tnote{$a$}  & 1500 &        0.5601  & 0.5607  & 1.2371  & 1.2303  & 0.2507  & 0.1611  & 1.7196  & 1.6794  \\
			SDEGen\tnote{$b$}  & 2000 &        0.5659  & 0.6358  & 1.2365  & 1.2246  & 0.2644  & 0.1731  & 1.7046  & 1.6997  \\
			SDEGen\tnote{$c$}  & 1500 &        0.2629  & 0.1015  & 1.6011  & 1.6030  & 0.1173  & 0.0302  & 2.0078  & 1.9853  \\
			SDEGen\tnote{$d$}  & 6000 &        0.6727  & 0.7420  & 1.1256  & 1.1289  & 0.3225  & 0.2565  & 1.6793  & 1.6587  \\
			GeoDiff\tnote{$e$}  & 1000  &        0.8296  & 0.9629  & 0.9525  & 0.9334  & 0.4827  & 0.4603  & 1.3205  & 1.2724  \\
			GeoDiff-A\tnote{$f$} & 5000  &      0.8836  & 0.9609  & 0.8704  & 0.8628  & 0.6014  & 0.6125  & 1.1864  & 1.1391  \\
			GeoDiff-C\tnote{$g$}  & 5000  &      \textbf{0.8913 } & \textbf{0.9788 } & \textbf{0.8629 } & \textbf{0.8529 } & \textbf{0.6147 } & \textbf{0.6455 } & \textbf{1.1712 } & \textbf{1.1232 } \\
			\midrule
			EcConf & 1     &       0.7972  & 0.8419  & 1.0233  & 0.9961  & 0.7049  & \textbf{0.8370 } & 1.1183  & 1.0597  \\
			EcConf & 2     &       0.7960  & 0.8333  & 1.0157  & 0.9995  & 0.7061  & 0.8297  & 1.1175  & 1.0583  \\
			EcConf & 5     &       0.8454  & 0.9118  & 0.9341  & 0.9264  & 0.7140  & 0.8317  & 1.0971  & 1.0270  \\
			EcConf & 10    &       0.8576  & 0.9228  & 0.9088  & 0.8975  & 0.7118  & 0.8205  & 1.0925  & 1.0243  \\
			EcConf & 15    &       0.8594  & 0.9255  & 0.9046  & 0.8905  & \textbf{0.7163 } & 0.8344  & \textbf{1.0841 } & \textbf{1.0176 } \\
			EcConf & 25    &       \textbf{0.8666 } & 0.9190  & \textbf{0.9016 } & \textbf{0.8869 } & 0.7136  & 0.8062  & 1.0931  & 1.0307  \\
			EcConf & \textbf{30} & 0.8640  & \textbf{0.9344 } & 0.9024  & 0.8993  & 0.7013  & 0.7945  & 1.1080  & 1.0489  \\
			\bottomrule
		\end{tabular}%
		
		\begin{tablenotes}
			\footnotesize
			\item$^a$SDE-Gen with default sampling settings: $n_{\text {euler }}=250, n_{\text {langevin }}=2, n_{\text {dsm }}=1000$.
			\item$^b$SDE-Gen with default sampling settings: $n_{\text {euler }}=500, n_{\text {langevin }}=2, n_{\text {dsm }}=1000$.
			\item$^c$SDE-Gen with default sampling settings: $n_{\text {euler }}=500, n_{\text {langevin }}=2, n_{\text {dsm }}=500$.
			\item$^d$SDE-Gen with default sampling settings: $n_{\text {euler }}=500, n_{\text {langevin }}=2, n_{\text {dsm }}=5000$.
			\item$^e$GeoDiff trains and samples with 1000 steps.
			\item$^f$GeoDiff-A trained with alignment approaches.
			\item$^g$GeoDiff-A trained with chain-rule approaches.
		\end{tablenotes}
	\end{threeparttable}
	
	\label{tab3}%
\end{table}%

\subsubsection{Performance of EC-Conf under various sampling iterations}
Once EC-Conf is trained, it allows either one step generation from the prior Gaussian distribution or carrying out iterative refinement via chaining the outputs of multiple time steps as shown in Algorithm \ref{algorithm2}. Here, we evaluate the performance of EC-Conf under various sampling iterations. In the meantime, rule based ETKDG method and 7 ML-based baselines were also compared, including one-shot methods: GraphDG, ConfVAE, GeoMol and iterative refinement methods: CGCF, ConfGF, SDEGen, and GeoDiff.

We evaluate our EC-Conf model on QM9 dataset, and the recall measurement of COV-R and MAT-R are as shown in Table \ref{tab2} of Appendix \ref{app1}, representing the diversity of generated conformations. It$'$s clear that the diffusion-based models out-performed one-shot models suggesting its better capability in reproducing ground truth conformations. Interestingly, one-shot generation of EC-Conf is already comparable with one-shot models, and the result for two iterations is almost the same as SDEGen model running 1500 iterations, i.e., sampling efficiency improved almost 750 times. The performance of the EC-Conf on the diversity gradually converged with more than 5 iterations. Although the recall performance is somewhat lower than GeoDiff model, the improvement on sampling efficiency of 1000 times could justify that EC-Conf model may be a better choice in dealing with large number of molecules. The precision-based metrics of COV-P and MAT-P are also as shown in Table \ref{tab2}  of Appendix \ref{app1}, where the EC-Conf out-performed all the baselines indicating much better quality of conformation generation under all sampling iterations. The optimal performance of EC-Conf was obtained at around 25 iterations.

\begin{figure*}[htbp]
	\centering
	\centerline{\includegraphics[width=5in]{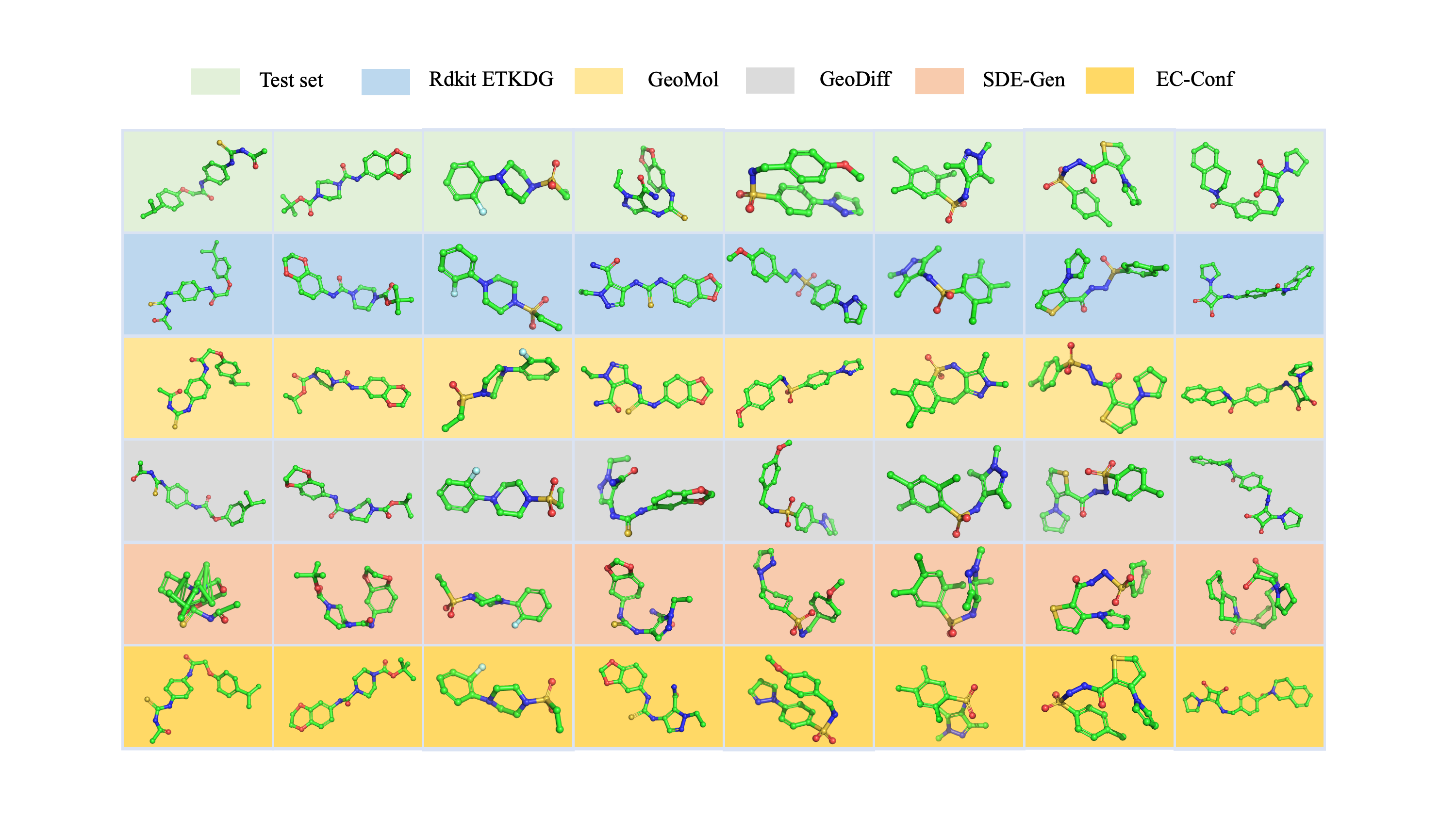}}
	\caption{Examples of generated structures for 8 random selected molecules in GEOM-Drugs test set with different methods.}
	\label{fig4}
\end{figure*}

We also evaluate our EC-Conf model on drug-like molecules with maximum of 50 heavy atoms in GEOM-Drugs sets. The recall-based metrics of COV-R and MAT-R are shown in Table \ref{tab3}. Again, the one-shot generation of EC-Conf greatly out-performed those one-shot baselines and some of the iterative methods such as CGCF, ConfGF, SDEGen. The performance of EC-Conf generation with 5 iterations already performed better than that of GeoDiff with 1000 iterations, representing 200 times better efficiency. The precision-based metrics on GEOM-Drugs set are also listed in Table \ref{tab3}. Although rule based RDKit method generally performed best, performance of EC-Conf model is quite close. Among all deep learning-based models, EC-Conf with single iteration out-performed other models and achieved best results with 15 sampling iterations. Some sample conformations generated by selected models are shown in Figure \ref{fig4} to provide a qualitative comparison, where EC-Conf is shown to nicely capture both local and global structures in 3D space.

\section{Conclusion}
In this paper, an equivariant consistency model (EC-Conf) was proposed as an ultra-fast diffusion method with only a few iterations for low-energy conformation generation. A time factor-controlled SE (3)-equivariant transformer was used to encode the Cartesian molecular conformations and a highly efficient consistency diffusion process was carried out to generate molecular conformations, largely achieving a balance between the conformation quality and computational efficiency. Our results demonstrate that EC-Conf can potentially learn the distribution of low energy molecular conformation with at least two magnitudes higher efficiency than conventional diffusion models and could potentially become a useful tool for conformation generation and sampling.

\bibliography{iclr2023_conference}
\bibliographystyle{iclr2023_conference}

\appendix
\section{Appendix} 
\label{app1}

\begin{table}[htbp]
	\centering
	\scriptsize
	\caption{Results of EC-Conf with different iteration steps on the GEMO-QM9 dataset}
	\begin{tabular}{c|cccc|cccc}
		\toprule
		\multirow{5}[6]{*}{\textbf{Steps}} & \multicolumn{4}{c|}{\textbf{QM9}} & \multicolumn{4}{c}{\textbf{QM9}} \\
		\cmidrule{2-9}          & \multicolumn{4}{c|}{\textbf{Conformation Generation}} & \multicolumn{4}{c}{\textbf{Conformation Generation}} \\
		\cmidrule{2-9}          & \multicolumn{2}{c}{COV-R $\uparrow$} & \multicolumn{2}{c|}{MAT-R $\downarrow$} & \multicolumn{2}{c}{COV-P  $\uparrow$} & \multicolumn{2}{c}{MAT-P $\downarrow$} \\
		& Mean  & Median & Mean  & Median & Mean  & Median & Mean  & Median \\
		&       &       &       &       &       &       &       &  \\
		\midrule
		5     & 0.4585  & 0.4298  & 0.5215  & 0.5202  & 0.8569  & 0.9228  & 0.3555  & 0.3513  \\
		10    & 0.7586  & 0.7977  & 0.3605  & 0.3521  & 0.8682  & 0.9090  & 0.3396  & 0.3368  \\
		15    & 0.7628  & 0.7907  & 0.3418  & 0.3383  & \textbf{0.8945 } & \textbf{0.9396 } & \textbf{0.3098 } & \textbf{0.3045 } \\
		25    & \textbf{0.8235 } & \textbf{0.8654 } & \textbf{0.3223 } & \textbf{0.3196 } & 0.8630  & 0.9088  & 0.3368  & 0.3356  \\
		50    & 0.8048  & 0.8571  & 0.3385  & 0.3307  & 0.8279  & 0.8712  & 0.3540  & 0.3513  \\
		\bottomrule
	\end{tabular}%
	\label{tab1}%
\end{table}%

\begin{table}[htbp]
	\centering
	\scriptsize
	\caption{The benchmark results on GEMO-QM9 test set}
	\begin{threeparttable}
		\begin{tabular}{l|c|cccc|cccc}
			\toprule
			\textbf{Dataset} &  &   \multicolumn{4}{c|}{\textbf{QM9}} & \multicolumn{4}{c}{\textbf{QM9}} \\
			\midrule
			\textbf{Task} &  &    \multicolumn{4}{c|}{\textbf{Conformation Generation}} & \multicolumn{4}{c}{\textbf{Conformation Generation}} \\
			\midrule
			\multicolumn{1}{l|}{\textbf{Metric} $\rightarrow$} & \multirow{2}[1]{*}{\textbf{Steps}} & \multicolumn{2}{c}{COV-R $\uparrow$} & \multicolumn{2}{c|}{MAT-R $\downarrow$} & \multicolumn{2}{c}{COV-P $\uparrow$} & \multicolumn{2}{c}{MAT-P $\downarrow$} \\
			\multicolumn{1}{l|}{\textbf{Model} $\downarrow$} &  &  Mean  & Median & Mean  & Median & Mean  & Median & Mean  & Median \\
			\midrule
			RDKIT & 1       &        \textbf{0.8326 } & \textbf{0.9078 } & \textbf{0.3447 } & \textbf{0.2935 } & $-$     & $-$     & $-$      & $-$  \\
			GraphDG & 1     &        0.7333  & 0.8421  & 0.4245  & 0.3973  & \textbf{0.4390 } & \textbf{0.3533 } & \textbf{0.5809 } & \textbf{0.5823 } \\
			ConfVAE & 1     &        0.7784  & 0. 8820 & 0.4154  & 0.3739  & 0.3802  & 0.3467  & 0.6215  & 0.6091  \\
			Geomol & 1      &        0.7126  & 0.7200  & 0.3731  & 0.3731  & $-$      & $-$      & $-$      & $-$  \\
			\midrule
			CGCF  & 1000        &    0.7805  & 0.8248  & 0.4219  & 0.3900  & 0.3649  & 0.3357  & 0.6615  & 0.6427  \\
			ConfGF & 5000       &    0.8849  & 0.9431  & 0.2673  & 0.2685  & 0.4643  & 0.4341  & 0.5224  & 0.5124  \\
			SDEGen\tnote{$a$}  & 1500       &    0.8153  & 0.8599  & 0.3568  & 0.3612  & 0.4837  & 0.4663  & 0.5662  & 0.5483  \\
			GeoDiff-A\tnote{$b$}  & 5000    &    \textbf{0.9054 } & \textbf{0.9461 } & 0.2104  & 0.2021  & 0.5235  & 0.5010  & 0.4539  & 0.4399  \\
			GeoDiff-C\tnote{$c$}  & 5000    &    0.9007  & 0.9339  & \textbf{0.2090 } & \textbf{0.1988 } & \textbf{0.5279 } & \textbf{0.5029 } & \textbf{0.4448 } & \textbf{0.4267 } \\
			\midrule
			EcConf & 1     &   0.7541  & 0.7935  & 0.4110  & 0.4085  & 0.7977  & 0.8252  & 0.4137  & 0.4148  \\
			EcConf & 2     &   0.7589  & 0.7971  & 0.4087  & 0.4016  & 0.7985  & 0.8285  & 0.4128  & 0.4176  \\
			EcConf & 5     &   0.8295  & 0.8806  & 0.3475  & 0.3440  & 0.8324  & 0.8766  & 0.3777  & 0.3732  \\
			EcConf & 10    &   \textbf{0.8314 } & \textbf{0.8811 } & 0.3308  & 0.3307  & 0.8507  & 0.8929  & 0.3570  & 0.3559  \\
			EcConf & 15    &   0.8298  & 0.8781  & 0.3253  & 0.3244  & 0.8579  & 0.9040  & 0.3476  & 0.3468  \\
			EcConf & 25    &   0.8235  & 0.8654  & \textbf{0.3223 } & \textbf{0.3196 } & \textbf{0.8630 } & \textbf{0.9088 } & 0.3368  & 0.3356  \\
			EcConf & 30    &   0.8131  & 0.8662  & 0.3241  & 0.3223  & 0.8614  & 0.9007  & \textbf{0.3330 } & \textbf{0.3332 } \\
			\bottomrule
		\end{tabular}%
		
		\begin{tablenotes}
			\footnotesize
			\item$^a$SDE-Gen with default sampling settings: $n_{\text {euler }}=250, n_{\text {langevin }}=2, n_{\text {dsm }}=1000$.
			\item$^b$GeoDiff-A trained with alignment approaches.
			\item$^c$GeoDiff-A trained with chain-rule approaches.
		\end{tablenotes}
	\end{threeparttable}
	
	\label{tab2}%
\end{table}%

\end{document}

%% file: math_commands.tex
\usepackage{amsmath,amsfonts,bm}

\def\eqref#1{equation~\ref{#1}}

\def\1{\bm{1}}

\DeclareMathAlphabet{\mathsfit}{\encodingdefault}{\sfdefault}{m}{sl}
\SetMathAlphabet{\mathsfit}{bold}{\encodingdefault}{\sfdefault}{bx}{n}

